\documentclass[prl,aps,twocolumn,superscriptaddress,longbibliography,noeprint]{revtex4-2}
  
\usepackage {graphicx}
\usepackage {amssymb}
\usepackage {amsmath}
\usepackage{color}
\usepackage{babel}

\newcommand{\reffig}[1]{Fig.~\ref{#1}}
\newcommand{\refeq}[1]{Eq.~(\ref{#1})}

\newcommand{\vect}[1]{\mathrm{\mathbf{#1}}} 

\newcommand{\bracketA}[3]{\ensuremath{\left\langle#1\right|#2\left| #3\right\rangle}}

\newcommand{\INVremark}[1]{}
 \newcommand{\INVtodo}[1]{}

\begin{document}

\title{High order Brunel harmonics and  supercontinuum formed by a weak optical pump in
  presence of a strong terahertz field}
 
\author{I. Babushkin}
\affiliation{Institute for Quantum Optics, Leibniz
  Universit\"at Hannover, Welfengarten 1, 30167 Hannover, Germany}
\affiliation{Cluster of Excellence PhoenixD (Photonics, Optics, and Engineering – Innovation Across Disciplines), Welfengarten 1, 30167 Hannover, Germany}
\affiliation{Max Born Institute, Max Born Str. 2a, 12489
 Berlin, Germany}
\author{A.~Demircan}
\affiliation{Institute for Quantum Optics, Leibniz
  Universit\"at Hannover, Welfengarten 1, 30167 Hannover, Germany}
\affiliation{Cluster of Excellence PhoenixD (Photonics, Optics, and Engineering – Innovation Across Disciplines), Welfengarten 1, 30167 Hannover, Germany}
\author{U.~Morgner}
\affiliation{Institute for Quantum Optics, Leibniz
  Universit\"at Hannover, Welfengarten 1, 30167 Hannover, Germany}
\affiliation{Cluster of Excellence PhoenixD (Photonics, Optics, and Engineering – Innovation Across Disciplines), Welfengarten 1, 30167 Hannover, Germany}
\author{A. Savel'ev}
\affiliation{Faculty of Physics, Lomonosov Moscow State University, Leninskie gory, 2 119991, Moscow, Russia}
\affiliation{Lebedev Physical Institute Russian Academy of Sciences, Leninskii prosp. 53, 119991, Moscow, Russia}

\begin{abstract}
  Brunel harmonics appear in the optical response of an atom in process of laser-induced ionization, when the electron leaves the atom and is accelerated in the strong optical field. 
  In contrast to recollision-based harmonics, the Brunel mechanism does not require the electron returning to the core.
  Here we show that in the presence of a strong ionizing terahertz (THz) field, 
  even a weak  driving field at the optical frequencies allow for generating  Brunel harmonics effectively.  
  The strong ionizing THz pump suppresses recollisions, making Brunel  dominant  in a wide spectral range. 
  High-order Brunel harmonics may form a coherent carrier-envelope-phase insensitive  supercontinuum,  compressible into an isolated
  pulse with the duration down to 100 attoseconds.
\end{abstract}

\maketitle

\textit{Introduction.} As it has been known from early stage of laser optics, interaction of
a strong optical field at frequency $\omega_0$ with matter creates
harmonics at $n\omega_0$, taking place via virtual bound-bound
atomic/molecular transitions \cite{boyd08:book}. 
Much later it was discovered that for higher intensities, when the atoms and molecules are ionized, high-harmonic generation (HHG) appears due to
the subsequent return of electrons to their parent cores
\cite{corkum89a}, leading to birth of attosecond science
\cite{corkum07} and allowing to break the femtosecond boundary for the pulse duration possessed by previous techniques
\cite{corkum94,corkum07}.

It was realized nearly at the same time that electron creation during ionization in a
strong field also leads to appearance of harmonics \cite{brunel90cp}
via bound-free and free-free transitions, that is, without need of the electron to come back. 
Here, following  
\cite{balciunas15,babushkin17} we will call the corresponding
nonlinearity ``Brunel nonlinearity'' and the arising harmonics will be
denoted as ``Brunel harmonics''. 
The ionization-induced nonlinearities
are typically ``nonperturbative'', that is, one can not obtain the
harmonics as series expansion in the vicinity of zero driver field --
in contrast to 
a typical case of nonresonant bound-bound
nonlinearities. 
This is related to the fact that Brunel harmonics are
proportional to the free electron density which, in turns, grows
exponentially fast with the field in the tunnel ionization regime \cite{keldysh1965ionization}. 

Brunel nonlinearity 
has been found to be useful, in
particular, for generation of low-frequency harmonics, in the terahertz (THz)
domain \cite{bartel05,reimann08}. The efficiency of such scheme for
THz generation can compete with 
optical rectification in
crystals \cite{yeh07,fueloep12}. 
More generally, generation of
high-amplitude THz radiation opens wide new perspectives \cite{dhillon17}, 
from remote detection for biological applications
\cite{reimann08} to compact electron accelerators
\cite{nanni15}.

Studies of the extreme-field THz interaction with matter are currently in their infancy due to quite a few available sources of extreme THz fields. 
Some work in this direction was also
done before, by exploring the influence of a ''strong'' DC or quasi-DC field on high harmonic generation  (HHG) 
by an optical field \cite{bao96hhg-static} and on generation of
attosecond pulses \cite{pan13unipolar}. 
This situation is changing
currently, in particular, due to novel sources of extreme THz fields exceeding 100~MV/cm
level \cite{vicario14,sell08},
allowing to consider processes appearing in the extreme-field regime at THz frequencies \cite{shkurinov17}, such as generation of high harmonics of the THz field \cite{schubert14}. 
In \cite{balogh11} it was shown that HHG from the intense mid-infrared (IR) field (670~MV/cm) can be controlled efficiently with a strong THz field (100 MV/cm at 33 THz). 
Here the optical field was high enough for tunnel ionization of Ne atoms, while the THz field modified (not too significantly) the ionized electrons paths, being not high enough to cause ionization.

 The limit, where the THz field can ionize atoms/molecules/solids directly, is of special interest since achievable THz field strengths exceed greatly the DC breaking threshold of a medium  \cite{chefonov17, chefonov18}. In particular, impact-ionization-induced transient optical nonlinearity was considered recently in \textit{p}-Si at THz field above 10 MV/cm \cite{savelev21}.

 In this paper, we consider HHG created by a weak  probe pulse at optical (in particular mid-IR) frequencies, in the presence of an extreme THz transient with intensity slightly below the onset of the tunnel ionization. 
 We show that  the weak optical probe  triggers tunnel ionization on the subcycle scale, leading to a strong nonlinear response of atoms governed by the Brunel mechanism. 
 We observe that 
the strong THz driver suppresses recollision-based
harmonics, making the Brunel nonlinearity dominant. 
When the optical probe has a few-cycle duration, the Brunel harmonics form a broad coherent supercontinuum with a flat spectral phase profile, independent on the carrier-envelope phase,  capable of compression to an isolated attosecond-scale pulse.

\begin{figure*}
\includegraphics[width=\textwidth]{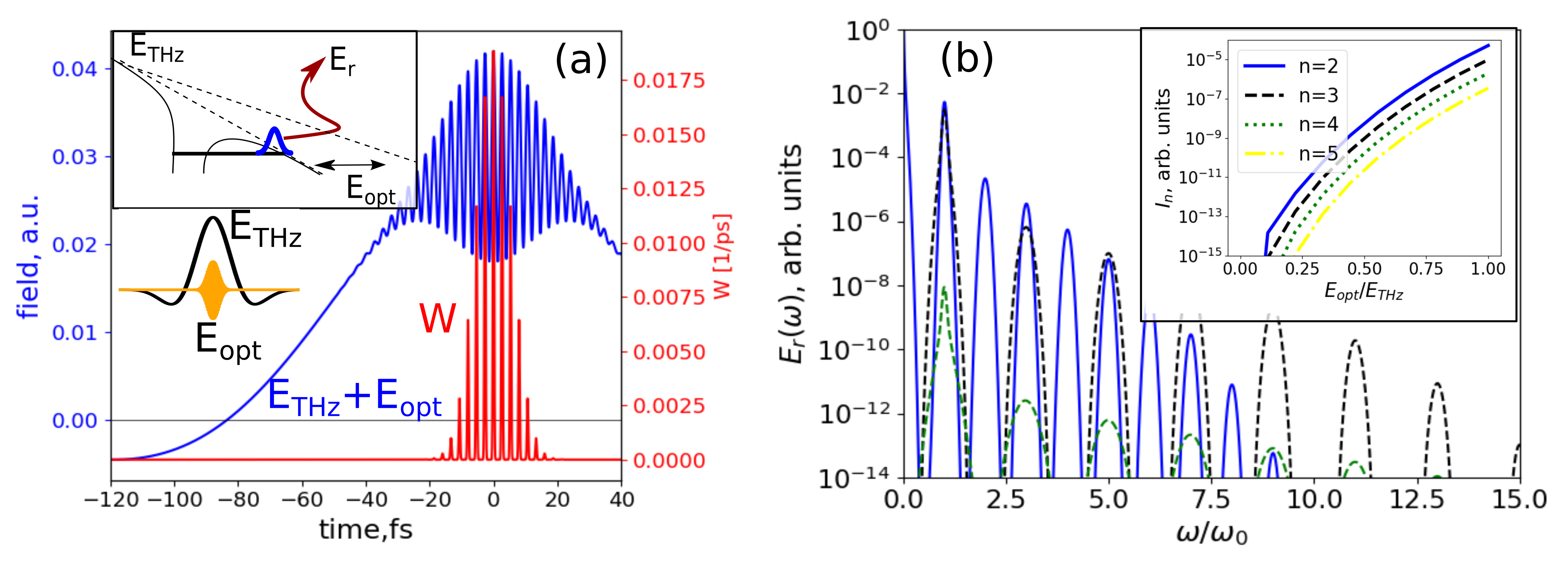}
\caption{ \label{fig:class}  Nonlinear response of argon to a weak optical probe pulse in
  presence of a strong THz field -- a simple-man picture. (a) Exemplary
  electric field (blue line) consisting of a strong THz
  ($E_\mathrm{THz}=0.03$~a.u.) centered at 100 $\mu$m wavelength and a
  weaker fundamental harmonic at 800 nm wavelength
  ($E_\mathrm{opt}=0.012$ a.u.), together with the ionization rate
  $W(t)$ (red
  line) induced by such a waveshape in argon according to the tunnel
  model. The lower inset shows the composition of the full driver signal from strong THz $E_{THz}$ and weak optical $E_{opt}$ field. The upper inset shows the schematics of the basic mechanism. (b) Corresponding Brunel harmonic response as given by
  \refeq{eq:drude} (blue line). For comparison, also the Brunel harmonics created by the optical field alone with the amplitude
  $E_\mathrm{opt}=0.042$~a.u. (corresponding to
  $E_\mathrm{opt}+E_\mathrm{THz}$ from (a), black dashed
  line)  and $E_\mathrm{opt}=0.012$~a.u. (corresponding to
  $E_\mathrm{opt}$ from (a), green dashed
  line). The latter is multiplied by $10^{35}$ for better visibility. The inset shows the intensity of $n$th  harmonics (for several values of $n$) in dependence on $E_{opt}$.}
\end{figure*}

\textit{Simple-man picture.} To start with, we formulate our idea using a simple-man quasi-static
tunnel-based model of the ionization in strong optical fields \cite{keldysh1965ionization}. In this
framework, the electron tunnels from the ground state to the continuum
through the barrier, created, on the one side, by the core potential,
and on the other side, by the strong external driving field (see \reffig{fig:class}a).
Effectively the electron can be considered as being born in continuum with ionization rate
$W(t) = (\alpha/|\vect E|) \exp{\left(-\beta/|\vect E|\right)}$, where
$\alpha$ and $\beta$ are some coefficients and $\vect E$ is the driving
field strength \cite{keldysh1965ionization}.
The free electrons form a current  \cite{kim09,babushkin10,babushkin17} given, under the assumption that the initial velocity is negligible, by (here and thereafter we use Hartree atomic units):
\begin{equation}
  \label{eq:drude}
  \frac{d \vect J}{dt} =  \rho(t)  \vect E(t).
\end{equation}
Here $W(t)$ and the free electron density $\rho(t)$ are connected by
$\frac{d\rho(t)}{dt} =W(t) (\rho_0(t)- \rho(t))$, where $\rho_0(t)$ is
the density of neutrals before the field is switched on. 

The change of the current, according to the Maxwell equations, serves as a source of electromagnetic radiation $\vect E_\mathrm{r} \propto
d\vect J/dt$, which is here refereed to as Brunel harmonics.   
When the initial driving pulse is a strong THz wave with intensity just below the onset of the tunnel ionization, the presence of even a weak pulse at the optical frequency, leads to a significant modulation of the tunneling probability at the subcycle scale, and thus to a strong nonlinear response (see inset to \reffig{fig:class}a). An exemplary simulation is shown in \reffig{fig:class},
where \reffig{fig:class}a shows the dynamics of the ionization of an argon atom for the case of the pump given by the linearly polarized THz field 
 of duration of 100 fs, central frequency 3 THz and
amplitude $E_\mathrm{THz}=0.03$ a.u. (156 MV/cm), 
whereas the  optical driver has a linear polarization, co-directed with the THz field, 
 with 10 fs duration and field amplitude $E_\mathrm{opt}=0.012$ a.u. (60 MV/cm), centered at 800 nm wavelength. In
\reffig{fig:class}b the corresponding spectrum is shown for this pump configuration by the blue line. As one can see by the  comparison to the case of the same $E_{opt}$ but zero $E_{THz}$ (green dashed line in \reffig{fig:class}b), the presence of a strong THz pump leads to a significant amplification of the Brunel nonlinearity and thus of the intensity of the Brunel harmonics. The inset to \reffig{fig:class}b also shows that this intensity grows quickly with increasing of $E_{opt}$.  
We see also that the
Brunel harmonics decay exponentially with the harmonics number, as it
is known also for other pump configurations \cite{babushkin11}.
Nevertheless, harmonics up to order $\sim$ 10 are yet quite
pronounced. 
In contrast to a single-pump configuration, both even and
odd harmonics are present. 
This is the obvious consequence of the fact
that the THz wave breaks the inversion symmetry of the whole system.
Also, for the sake of comparison, in \reffig{fig:class}b (black dashed line) we plotted
Brunel harmonics given by the purely optical field
$E_\mathrm{opt}=0.042$ a.u., corresponding to $E_{THz}+E_{opt}$ in the previous configuration. One can see that also for the odd harmonics of relatively low
order (below $n\sim5$), our two-color configuration
provides an advantage in efficiency over the single-color one. 
As we see in the next session by considering a more detailed picture, Brunel harmonics in the single color configuration have no advantage also for higher frequencies $n\gtrsim 5$.

\begin{figure*}
\includegraphics[width=\textwidth]{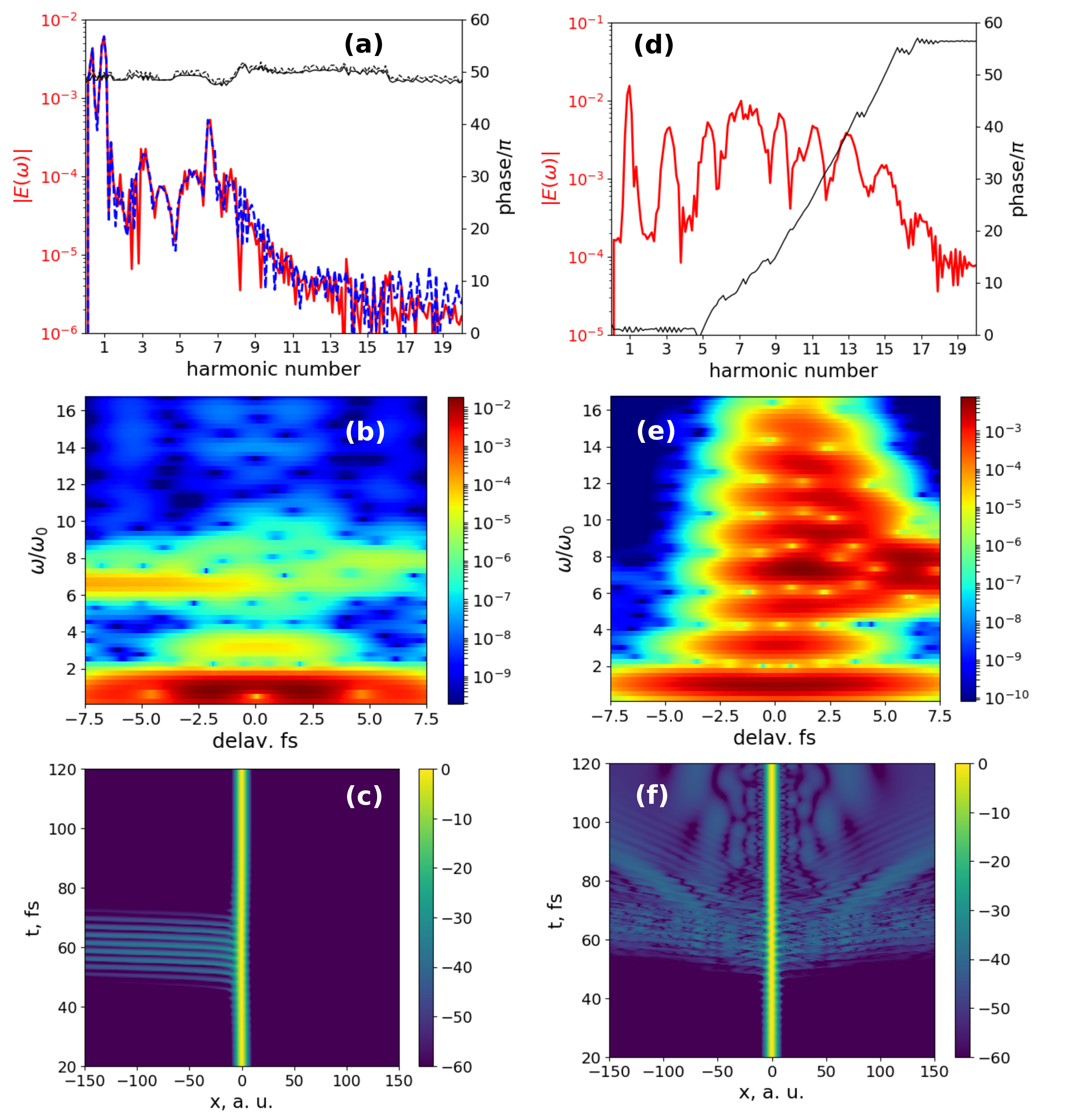}
\caption{ \label{fig:3d} The response of the hydrogen atom according to TDSE simulations [\refeq{eq:1a}] with $E_{opt}$ given by a 5 fs-long pulse
  with (a,b,c) and without (e,d,f) a strong THz field. The field amplitudes
  are as in \reffig{fig:class}. (a,c) The harmonic spectrum of the
  atomic response $\vect E_r$ according to \refeq{eq:br-q}. Red and blue lines: spectral amplitudes, black lines: spectral phases. In (a), the cases of the CEO phase $=0$ (solid lines) and $\pi$ (dashed lines) are shown.  (b,d) Corresponding XFROG
  traces. (e,f) Spatio-temporal dynamics of the electron wavepacket
  $|\psi(x,t)|^2$ simulated by 1D-variant of \refeq{eq:1a} with a regularized Coulomb potential (see text).  }
\end{figure*}

\textit{Quantum-mechanical approach.} As a next step, we simulate the ionization process using 
the time-dependent Schr\"odinger equation (TDSE) for the hydrogen atom: 
\begin{equation}
    i \partial_t \psi(\vect r,t) = H\psi(\vect r,t), \, H=(\vect
    p+\vect A(t))^2/2+V(r), 
    \label{eq:1a}
\end{equation}
where $\psi(\vect r, t)$ is the wavefunction of the electron depending
on space $\vect r$ (with $r\equiv |\vect r|$) and time $t$ coordinates, $H$ is the
electron's Hamiltonian, with $\vect p$ being the momentum operator,
$\vect A(t)$ the vector potential of the driving field corresponding to 
the electric field strength $\vect E(t) = -\partial_t \vect A$, and
$V=-1/r$ is the potential created by the hydrogen core. The optical
response of the atom, which in this case includes not only Brunel
mechanism but also bound-bound transitions and recollision harmonics, is given by
\begin{equation}
  \label{eq:br-q}
  \vect E_r = \frac{d^2}{dt^2} 
  \bracketA{\psi}{\vect r}{\psi}.
\end{equation}
The
code developed in \cite{patchkovskii16} was used to perform the numerical integration, with the simulation box size of 500 a.u.. The driving field was given
by $\vect A = \vect A_\mathrm{THz} + \vect A_\mathrm{opt}$,   
\begin{equation}
  \label{eq:Aopt}
  A_{x,\mathrm{opt}}= e^{-\frac{t^2}{\tau_\mathrm{opt}^2}}\sin{(\omega_0 t)}, \, A_{x,\mathrm{THz}}= e^{-\frac{t^2}{\tau_\mathrm{THz}^2}}\sin{(\omega_\mathrm{THz} t)}, 
\end{equation}
and  $A_{y,\mathrm{opt}}=0$, $A_{y,\mathrm{THz}}=0$, 
where $\omega_0$ is frequency  corresponding to 800 nm wavelength
and $\omega_\mathrm{THz}=3\times2\pi$ THz, $\tau_\mathrm{opt}$
corresponds to 5 fs full-width half-maximum (FWHM)  and $\tau_\mathrm{THz}$ corresponding to 100
fs FWHM pulse durations. 

The simulation results are shown in \reffig{fig:3d}a,b for $E_\mathrm{THz}$ and $E_\mathrm{opt}$ the same as in \reffig{fig:class}a. In
\reffig{fig:3d}a, the spectrum of $\vect E_\mathrm{r}$ (for two different carrier-envelope (CEO) phases, 0 and $\pi$,  and in (b) the XFROG trace for the zero CEO phase is presented.
For comparison, in \reffig{fig:3d}c,d the same pictures are shown for the strong optical pump only (equal to the sum of $E_{opt}$ and $E_{THz}$ from the previous case). As one can see, in the latter case high order harmonics up to $\sim$ 17 appear. Examining the XFROG \reffig{fig:3d}d we see that these harmonics originate from the recollision harmonics. Indications of this is the characteristic time-dependent delay in harmonic emission visible in \reffig{fig:3d}d. 
In contrast, in the former case of THz+optical pump
the harmonics have
no systematic delay,
 pointing to the Brunel generation mechanism.  We see
also that the harmonics are 
less localized in time in the two-color case,
which is also good explained by the Brunel mechanism, since
 Brunel harmonics are born in the process of both electron birth and
acceleration, so their generation 
should take certain time.
The different generation mechanism is furthermore indicated by radically different spectral phases for two- and single-color cases, shown by the black lines in \reffig{fig:3d}a and \reffig{fig:3d}d. The former is almost frequency independent, in strong contrast to the latter. 
The spectrum in the two-color case is extended only up to around 10th harmonic
in a qualitative agreement with the simple-man theory (see
\reffig{fig:class}b),
also supporting the Brunel mechanism for that case.

This suggests, that recollision-based harmonics are significantly suppressed. 
To obtain an additional confirmation of this suppression,
we made simulations of one-dimensional variant of \refeq{eq:1a} 
with a soft-core Coulomb
potential $V=-1/\sqrt{x^2+a^2}$, which, for $a=1/\sqrt{2}$ has the same  ionization potential as the hydrogen atom \cite{eberly89,hall09}. The results of simulation for the same parameters of as in \reffig{fig:3d}a and \reffig{fig:3d}d are
given in \reffig{fig:3d}c and \reffig{fig:3d}f respectively.
The advantage of the
one-dimensional simulation is that one can very clearly see the trajectories of
the electrons: If the strong THz field is present (\reffig{fig:3d}c), the electrons 
move from the core without making even a single recollision, in
a strong contrast to the case of the  single color optical pump (\reffig{fig:3d}f). This is explained by the fact that 
the ponderomotive energy gained by the
electron from the THz field over the half of the optical field cycle  
$\propto E_\mathrm{THz}^2$ 
which is in our case one order of magnitude larger than the 
energy gained form the optical field  $\propto E_\mathrm{opt}^2$ over the same time. Because the subsequent optical half-cycles are located within the very same THz cycle, the electron do have no chances to return back. 

\begin{figure}
\includegraphics[width=\columnwidth]{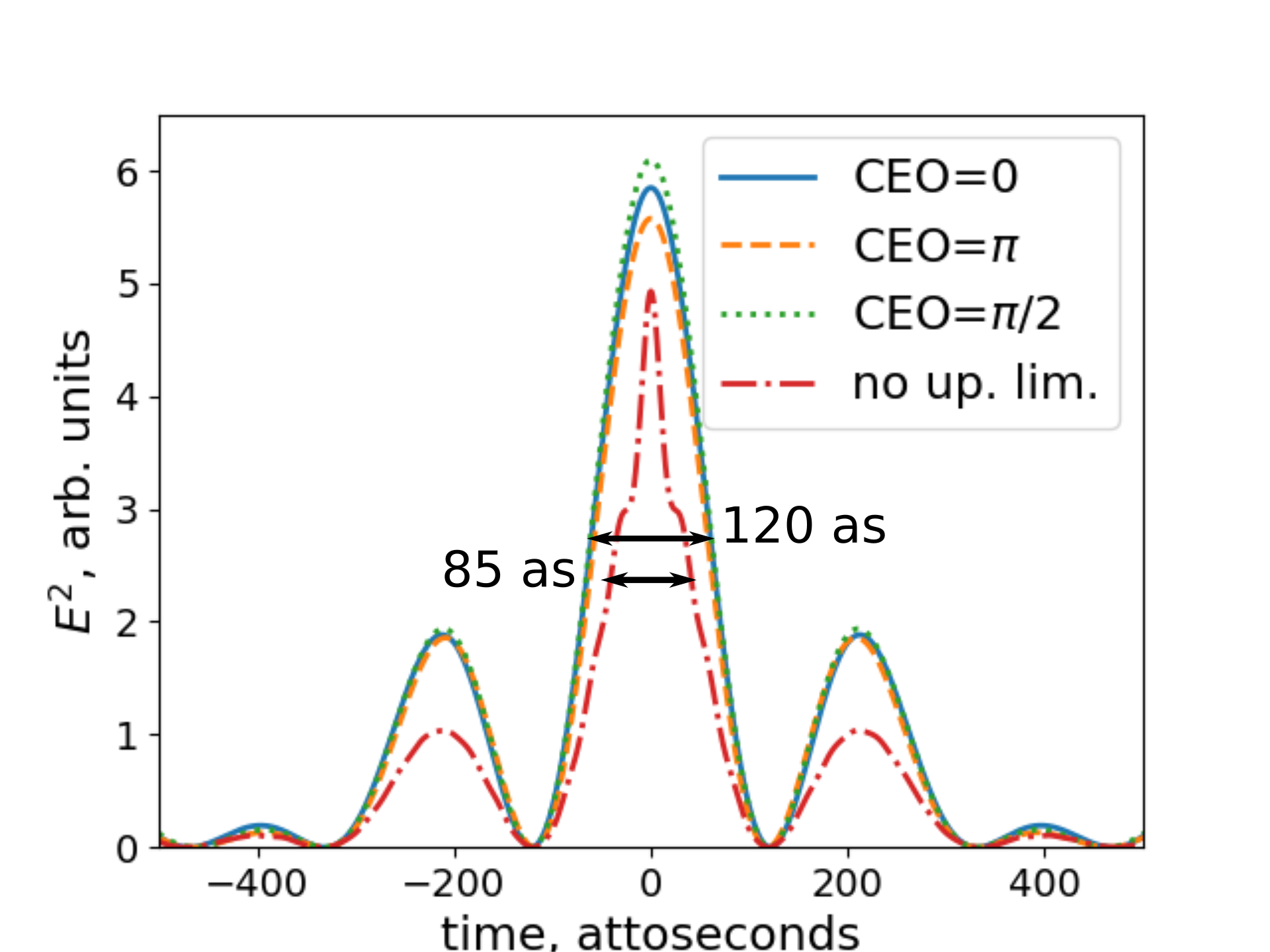}
\caption{ \label{fig:compr}  Solid blue line: The square of the electric field in dependence on time for the Fourier-limited pulse obtained from the spectrum  spectrum in in \reffig{fig:3d}a in the range between  $\omega=1.5\omega_0$ and $\omega=10\omega_0$. FWHM duration of the resulting pulse is 120 attoseconds. To demonstrate the CEO-insinsitivity, dashed orange and dotted green lines show the cases of the different CEO phase of the optical driver, compressed with the phase mask for the zero CEO phase. Red dash-dotted line shows the compressed pulse with the same lower but with no upper filtering limit. In this case FWHM duration is 85 fs.   
}
\end{figure}

Importantly, the spectrum in \reffig{fig:3d}a forms a broad continuum 
in the
range of $n\approx 1.5-10$ harmonics. The most important reason why  separate harmonics join into a broad continuum is rather the
short optical probe, leading to broadening of every harmonic; Besides, in our case both odd and even
harmonics are present, so the distance between the subsequent harmonics is two times smaller than in a single-color case.
The spectral phase of the resulting continuum is very flat across the whole spectral range (see black line in \reffig{fig:3d}a);
Removing this phase leads to a stand-alone Fourier-limited  compressed pulse of a 100-attosecond duration as shown in \reffig{fig:compr}. If the spectral range from 1.5th to 10th harmonic is taken for compression, the resulting pulse is  120 attoseconds long, whereas if the whole  spectrum above  1.5th harmonic is used, the final pulse duration is reduced to 85 attoseconds. 

Very importantly, both the spectral intensity and the spectral phase are rather insensitive to the CEO phase of the optical probe (see \reffig{fig:3d}a, where the cases of different CEO phases are shown by solid and dashed lines).  This is in a strong contrast to the continua created by recollision-based harmonics (which are also located at higher frequencies) \cite{christov97,brabec00,chini14-att-rev}. These continua  are heavily CEO-dependent since they rely on the inter-cycle electrons trajectories leading back to the core. 
In contrast, the Brunel radiation is emitted during the ionization event, taking place on the subcycle scales. Interestingly, the CEO-insensitivity of the continuum is also rather different from the behaviour of separate harmonics obtained from longer pump pulses (for instance in \reffig{fig:class}b) where the CEO phase of $n$th harmonic is $n$ times the CEO phase of the fundamental.  
This pronounced CEO-insensitivity leads also to the insensitivity of the  resulting compressed pulse to the CEO phase, as shown in \reffig{fig:compr} where compression of spectra resulted from  optical probe with different CEO phases is made using the same phase mask. That is, the compression scheme does not require CEO stabilization of the optical probe. As we see, these pulses do not require also gating necessary for isolating attosecond pulses for recollision-based harmonics \cite{christov97,brabec00,chini14-att-rev}.  

\textit{Conclusion and discussions.} In conclusion, we showed that in the presence of a strong ionizing THz driver, 
higher order harmonics can be generated by a weak optical probe pulse via the
Brunel (ionization-induced) mechanism. 
This mechanism dominates over recollision-based harmonics,  
since the recollisions 
are  strongly 
suppressed in the presence of the strong THz wave. 
When generated with a few-cycle optical pump, the harmonics in the range from visible to XUV 
may form a broadband supercontinuum, compressible to an isolated 100-attosecond-scale pulse without need for gating techniques. The generation efficiency is comparable to the one based on recollision harmonics, but taking into account that the spatial area of focused THz pulses is around four orders of magnitude larger than the area at optical frequencies, and, since only a weak optical driver is needed (which thus can be significantly defocused), the overall energy yield can be several orders of magnitude higher than from the  recollision-based mechanism.      

Importantly, this approach is fully insensitive to the CEO phase of the optical driver, so that CEO stabilization  is not required.   
CEO-phase-insensitivity implies also shot-to-shot coherence of the supercontinuum, the property which is difficult to achieve using traditional fiber-based supercontinuum generation methods \cite{demircan05,dudley06,babushkin17a}. 

\begin{acknowledgments}
A.S. acknowledges support from the Russian Science Foundation under project 20-19-00148. I.B. and U.M. 
thank Deutsche Forschungsgemeinschaft (DFG, German Research
Foundation), projects BA 4156/4-2 and MO 850-19/2 for support. I.B., A.D. and U.M. acknowledge support from 
Germany’s Excellence Strategy within the Cluster of Excellence
PhoenixD (EXC 2122, Project ID 390833453).  
\end{acknowledgments}


%

\end{document}